\documentstyle{cupconf}
\include{epsf}
\ifoldfss
\else
  \ifnfssone
    \newmathalphabet{\mathit}
      \addtoversion{normal}{\mathit}{cmr}{m}{it}
      \addtoversion{bold}{\mathit}{cmr}{bx}{it}
    \newmathalphabet{\mathcal}
      \addtoversion{normal}{\mathcal}{cmsy}{m}{n}
    \else
    \ifnfsstwo
    \fi
  \fi
\fi

\def\hmpc{\ifmmode{h^{-1}\,\hbox{Mpc}}\else{$h^{-1}$\thinspace Mpc}\fi}
\def\hkpc{\ifmmode{h^{-1}\,\hbox{kpc}}\else{$h^{-1}$\thinspace kpc}\fi}

\def\kms{\ifmmode{\,\hbox{km}\,s^{-1}}\else {\rm\,km\,s$^{-1}$}\fi}
\def\kpc{{\rm\,kpc}}
\def\mpc{{\rm\,Mpc}}
\def\msun{{\rm\,M_\odot}}

\def\cm{{\rm\,cm}}
\def\yr{{\rm\,yr}}
\def\gyr{{\rm\,Gyr}}

\def\gm{{\rm\,g}}
\def\psec{{\rm\,s^{-1}}}
\def\kmsm{{\rm\,km\,s^{-1}\,Mpc^{-1}}}

\def\spose#1{\hbox to 0pt{#1\hss}}
\def\lta{\mathrel{\spose{\lower 3pt\hbox{$\mathchar"218$}}
     \raise 2.0pt\hbox{$\mathchar"13C$}}}
\def\gta{\mathrel{\spose{\lower 3pt\hbox{$\mathchar"218$}}
     \raise 2.0pt\hbox{$\mathchar"13E$}}}

\def\hexnumber#1{\ifcase#1 0\or1\or2\or3\or4\or5\or6\or7\or8\or9\or
 A\or B\or C\or D\or E\or F\fi }

\makeatletter
\ifx\CUP@mtlplain@loaded\undefined
\else
\fi
\makeatother
 \makeatletter
 \ifx\CUP@mtlplain@loaded\undefined
   \font\tenbmi=cmmib10 at 10pt
   \font\sevenbmi=cmmib10 at 7pt
   \font\fivebmi=cmmib10 at 5pt

   \newfam\bmifam
   \textfont\bmifam=\tenbmi
   \scriptfont\bmifam=\sevenbmi
   \scriptscriptfont\bmifam=\fivebmi
   
 \fi
 \makeatother
\ifnfsstwo

\fi
\ifnfssone

\fi
\ifoldfss

\fi

\mathchardef\varLambda="0103

\makeatletter
\ifx\CUP@mtlplain@loaded\undefined
\else
\fi
\makeatother
\makeatletter
\ifx\CUP@mtlplain@loaded\undefined
  \font\tenbms=cmbsy10
  \font\sevenbms=cmbsy10 at 7pt
  \font\fivebms=cmbsy10 at 5pt
  \newfam\bmsfam
  \textfont\bmsfam=\tenbms
  \scriptfont\bmsfam=\sevenbms
  \scriptscriptfont\bmsfam=\fivebms

  \edef\bsy@{\hexnumber\bmsfam}
  \mathchardef\bnabla="0\bsy@72
\fi
\makeatother
\def\eg{{e.g.\ }}

\def\etal{\mbox{\it et al.}}

\title[Bulge Building with Mergers and Winds]{
Bulge Building with Mergers and Winds}

\author[R. G. Carlberg]
{R.\ns G.\ns C\ls A\ls R\ls L\ls B\ls E\ls R\ls G}
\affiliation{Department of Astronomy, University of Toronto,
Toronto ON, M5S 3H8, Canada}

\begin{document}
\ifnfssone
\else
  \ifnfsstwo
  \else
    \ifoldfss
      \let\mathcal\cal
      \let\mathrm\rm
      \let\mathsf\sf
    \fi
  \fi
\fi

\maketitle

\begin{abstract}
The gravitational clustering hierarchy and dissipative gas processes
are both involved in the formation of bulges. Here we present a simple
empirical model in which bulge material is assembled via gravitational
accretion of the visible companion galaxies.  Assuming that merging
leads to a starburst, we show that the resulting winds can be strong
enough that they self-regulate the accretion.  A quasi-equilibrium
accretion process naturally leads to the Kormendy relation between
bulge density and size.  Whether or not the winds are sufficiently
strong and long lived to create the quasi-equilibrium must be tested
with observations.  To illustrate the model we use it to predict
representative parameter dependent star formation histories.  We find
that bulge building activity peaks around redshift two, with tails to
both higher and lower redshifts.
\end{abstract}

\firstsection 

\section{Introduction}

Bulges are stellar dynamical pressure supported systems that generally
have much higher surface brightnesses than galactic disks. They
therefore have undergone more collapse than galactic disks, evidently
with the angular momentum barrier removed.  Galaxy merging is an
inevitable process that redistributes any pre-merger stars into a
physically dense, but phase density lowered, pressure supported
distribution. Stellar dynamical mergers produce objects with
flattenings largely unrelated to their rotation. In the presence of
gas, merging is empirically associated with an often dramatic rise in
star formation. These new stars that are formed in place almost
certainly reflect the chemical history and the dynamical state
of the growing bulge.  This paper calculates some of the properties of
bulges expected on the basis of merging with star formation of largely
gaseous pre-galactic fragments.

The rate of major mergers can be calculated directly from the observed
numbers of close pairs of galaxies. Remarkably, this is now an observational
quantity for which there are measures from low redshift to the
``U-dropout'' population centred around redshift three.  There are
some substantial uncertainties in the various observational quantities
which go into the merger calculation.  The details of this calculation
will become much more precise over the next few years as the evolution
of the two-point correlation function becomes better determined.

Mergers are widely observed to induce an intense nuclear starburst.
Theoretically this is at least partially understood (\cite{bh_rev}) on
the basis that the strong dynamical interactions during a merger leads
to a loss of angular momentum in a cool gas, helping to funnel it to
build up a dense central gas reservoir from which stars form at
astonishing rates in a starburst. Starbursts in turn develop winds
which we suggest can lead to the accretion being a self-regulating,
although this is dependent on the ram pressure and duration of the
wind. Moreover, self-regulating accretion can lead to
quasi-equilibrium star formation in the bulge which can lead to some
of the observed regularities of bulge properties with mass or size. At
this stage the details of this picture are speculative, but are open
to observational refinement, which helps motivate the calculations
presented here.

Other papers presented at this meeting describe in detail the
properties of bulges. Here we take the properties of ``classical
bulges'' to be roughly as follows (\eg\ \cite{wgf}).
\begin{itemize}
\item Bulges follow the ``Kormendy relations'', that is, the
	characteristic surface brightness correlate strongly with the
	scale radius (\cite{k77,dejong,pdd}).
\item The flattening of the figure of bulges is approximately
	consistent with their rotation (\cite{defis}).
\item Bulge stellar populations are predominantly old, although
	there are well-documented cases of relatively young
	bulges.
\item Bulges have a mass-metallicity correlation.
\end{itemize}
A useful model for bulges should be able to provide a physical origin
for these properties.

The paper has three main sections. In \S2 we discuss the empirically
determined rate at which mergers occur as a function of redshift.
Then, in \S3, we discuss star burst winds, and the effects those winds
will have on accreting gas. Section 4 pulls these two together in some
specific model calculations.

\section{Merger Rate Measures}

A host galaxy has a number of near neighbors within radius $r$ and
pairwise velocity $|v|$ far above the mean density $n_0$ (in proper
co-ordinates),
\begin{equation}
N(<r,<v)= 4\pi n_0(1+z)^3 \int_0^v\int_0^r \xi(r|z) f(v|z) r^2v^2\, dr\,dv, 
\end{equation}
where $\xi(r|z)$ and $f(v|z)$ are the redshift dependent two point
correlation and velocity distribution functions, respectively.  We
have made the important assumption that the distribution of pairwise
velocities is constant over the separations of interest. This is not
true in general, but is sufficient for our application to the
relatively small scales, $r\le 20 \kpc$, that are of interest for
merging. To calculate merger rates we need estimates of the
correlation function on small scales, the pairwise velocity
dispersions, and the mean time for a merger to occur within this
volume.

\subsection{Close Pairs and the Correlation Function}

The galaxy-galaxy correlation function is accurately modeled as a
power law, $\xi(r)=(r_0/r)^\gamma$. The reliability of this power law
on scales less than about 100\hkpc\ relevant to galaxy merging is
discussed in detail for the SSRS2 (\cite{patton_lowz}) and CNOC2
samples (\cite{patton_cnoc}). These papers support three important
conclusions. First, the power law extrapolation of the correlation
function to 20\hkpc\ is consistent with the density of pairs measured
inside this radius. Second, the R band luminosity function of galaxies
in 20\hkpc\ pairs is consistent with being drawn from the field
luminosity function. This property allows the fully general
correlation function, $\xi(r,v_{||},v_{\perp},L_1,L_2)$, to be
factored into a luminosity function, and a spatial and kinematic
correlation function, $\phi(L_1)\phi(L_2)\xi(r) f(v)$ where we drop
the distinction between velocities along the line of separation,
$v_{||}$, and perpendicular velocities, $v_\perp$ (\cite{lss}). The
luminosity factorization glosses over the various lines of evidence
(\cite{roysoc,loveday}) that the theoretically expected weak increase
of correlations with galaxy mass does exist in the correlations.
However, this relatively small effect cannot yet be detected in the
current samples of close pairs which have not yet broken through the
barrier of 100 pairs. The third important result is that there is
morphological evidence that $r\le 20\hkpc$ pairs are indeed
interacting at a level that indicates that these are high probability
mergers-to-come.  The 20\hkpc\ scale is also chosen such that the
galaxies are not so strongly interacting that their luminosities,
morphologies and colors bear little resemblance to their unperturbed
values.

\begin{figure}
\centering
\vspace*{5pt}
\parbox{\textwidth}{\epsfysize=7truecm\epsfbox{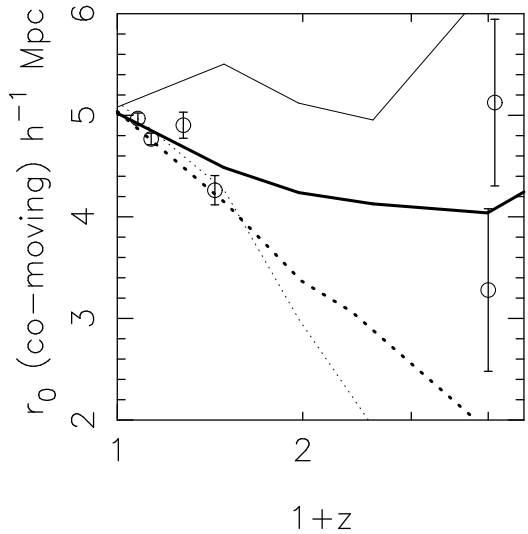}}
\vspace*{5pt}
\caption{Measured co-moving correlation lengths as a function 
	of redshift. The points are the LCRS at low redshift, CNOC2 at
	intermediate redshift, and the Giavalisco \etal\ (1998) close
	pairs and the ``box counts'' of Adelberger \etal\ (1998) at
	redshift 3. The lines are for the correlation lengths measured
	in simulations (\cite{ccc}) for galaxy halos (solid lines)
	and the mass field (dashed), with $\Omega_M=0.2$,
	and $\Omega_\Lambda=0$ (heavy) or $\Omega_\Lambda=0.8$ (light).
\label{fig:r0_z}}
\end{figure}

The volume integral of the power law correlation function,
$\xi(r|z)=(r_0(z)/r)^\gamma$, is,
\begin{equation}
4\pi \int_0^r \xi(r|z) r^2\, dr =
        {{4\pi}\over{3-\gamma}} \left( {r_0(z)\over r}\right)^\gamma r^3.
\label{eq:cor20}
\end{equation}
The redshift dependence of the average density inside a $r=20\hkpc$
neighborhood around a galaxy is estimated using the preliminary CNOC2
correlation $\gamma=1.8$, 
\begin{equation}
r_0(z)=5.15(1+z)^{[1-(3+\epsilon)/\gamma]}\hmpc\ \hbox{(co-moving)},
\end{equation}
where $\epsilon=-0.6\pm 0.4$ (\cite{roysoc}).  Using this in
Equation~\ref{eq:cor20} we find that the integrated density inside
20\hkpc\ is $1.56 [(1+z)/1.3]^{-\epsilon}n_0 \mpc^3$ (proper
units). In Figure~\ref{fig:r0_z} we show the co-moving correlation
length as a function of redshift. We also plot the correlations of the
galaxy mass halos and the particles in simulations (\cite{ccc}). It
should be noted that there is a relatively good understanding of why
the halo correlation function shows relatively little evolution
(\cite{cco,baugh}).  Observationally it is currently acceptable to
take $r_0$ to be fixed over this redshift interval, or,
$\epsilon=-1.2$ for a $\gamma=1.8$ power law.

\subsection{Pairwise Velocities}

The CNOC2 velocity distribution function, $f(v)$, at small scales is
acceptably modeled as a Gaussian, although an exponential also
provides a similar quality fit (\cite{roysoc}).  We will take the
velocity to be isotropic, $\sigma_3=\sqrt{3}\sigma_z$, where
$\sigma_z$ is the velocity dispersion measured along the line of
sight.  Redshift surveys can be used to measure $\sigma_z$. At
separations of about 1 \hmpc\ the the pairwise velocity dispersion in
the redshift direction is about 300\kms\ (\eg\
\cite{dp,marzke,roysoc}), constant with redshift over the $z \le 0.5$
range. If the critical velocity to merge is taken as about
$220\sqrt{2}$ \kms\ then the fraction of close pairs with sufficiently
low velocities to merge is merely 5.1\%. This does not accord well
with the impression that most close pairs have such large tidal tails
that they are almost certainly doomed-to-merge pairs (\cite{tt,t77}).
In a non-merging, equilibrium distribution the pairwise velocity
declines as $\sigma_p^2\propto r^{2-\gamma}$, where $\gamma$ is the
slope of the power law correlation, The one dimensional pairwise
velocity dispersion at 20\kpc\ is therefore about 190\kms.

The dynamical details of pair mergers in a cosmological setting that
includes the tidal fields of surrounding structure have not been
studied in detail (but see
\cite{cc89}), such that one of the best estimates of the critical
velocity to merge remains Aarseth and Fall's (1980) value of
$v_{mg}=1.2\sqrt{2} v_p$, where $v_p$ is the ``parabolic'' velocity,
$v_p=f_pv_c$, at the orbital pericenter, where $f_p$ is at least
$\sqrt{2}$ (for a point mass).  This ``parabolic'' velocity is that
for those pairs assured to merge, in the absence of tidal fields
(\cite{tt,af}). For $f_p=\sqrt{2}$ the fraction of the the close pairs
with velocities low enough to merge is 27\% but this rises quickly
with increasing $f_p$, becoming 41\% and 54\% at $f_p=\sqrt{3}$ and 2,
respectively. We will use an $f_{mg}=0.5$ but recognize that this
number is both empirically and theoretically uncertain.

\subsection{Merger Times of Close Pairs}

As a reference timescale for merging we start from our reference
radius of 20 kpc where the time for a circular orbit is 0.62 Gyr at a
speed of 220\kms. In detail the rate of inflow through the 20 kpc
depends on mass and the orbital details, so we use a merger timescale
at our reference radius of 20 kpc of 0.3 Gyr (\cite{bh_rev,dubinski}).
We use our estimate that $f_{mg}=0.5$ of all 20\kpc\ pairs have
pairwise velocities less than the critical velocity for merging. If
the pairwise velocities and substantially higher than the critical
velocity for merging then the merging fraction drops nearly as $v^3$,
which is such a dramatic change that it should be testable via the
morphology of galaxies in close pairs.  
\subsection{Estimated Merger Rates}

Combining our estimates of clustering, pairwise velocities and the
available understanding of the dynamics of merging, we find that the
specific merger rate is $2.4[(1+z)/1.3]^{-\epsilon}n_0 \mpc^3
\gyr^{-1}$.  If we take $n_0$ as being the CNOC2 galaxy number density
to $0.21L_\ast$ adjusted to $0.1L_\ast$, $n_0\simeq
1.1\times10^{-2}\mpc^{-3}$ (co-moving) the merging event rate for
galaxies above our minimum mass is $1.3\times 10^{-2}
(1+z)^{-\epsilon} \gyr^{-1}$.  The rate of accretion of pre-merger
stars onto galaxies as a result of merging is relatively slow. The
time scale is $L_\ast/{\cal R}_L\simeq 60 \gyr$, or about 5 Hubble
times at $z=0.3$. This result is based on the directly visible
galaxies, $L\gta 0.21 L_\ast$, which contain about 80\% of the total
stellar luminosity galaxies.  This relatively low rate of accretion of
visible galaxies relative to the Hubble times continues on to the
$z\simeq 3$ regime (\cite{mauro,adelberger}), where the number
densities and the co-moving correlations are similar to those observed
for present day galaxies. That is, for $q_0=0.1$ their data indicate a
density of ${\cal R} <25.5$ mag Lyman break galaxies of $n_0\simeq
2.2\times 10^{-3} \mpc^{-3}$ and a co-moving correlation length of
$5\hmpc$. The self-event rate of this population is $2.0\times^{-2}
\gyr^{-1}$, only 3.8 times that at $z\simeq 0.3$.  Since the cosmic
time at $z\simeq 3$ is about 20\% of that at $z=0.3$ the relative
impact on the hosts of these self-mergers is small adding perhaps
5-10\% more mass over the entire $z=0$ to 3 range.  For the lower
luminosity galaxies (about 2 magnitudes fainter than those with
spectroscopic redshifts) inferred to be in this redshift range from
the HDF the volume density is about a factor of 20 higher, but the
implied cross-correlation length is about a factor is about a factor
of $\sqrt{3}$ smaller (\cite{steidel}), where we assume that the
cross-correlation depends on the product of the relative biases. This
implies that the high-low luminosity merger rate is about $0.5
\gyr^{-1}$ which is large enough to build a galaxy
over a $z\simeq 1-4$ interval.  In the intermediate redshift range the
$M\gta0.2 M_\ast$ galaxies cannot self-merge to significantly alter
the mass function. At higher redshift the lower luminosity Lyman break
galaxies rise very steeply in number, $\alpha\simeq-1.8$
(\cite{steidel}). These large numbers completely change the situation,
allowing their mergers to substantially alter galaxy masses. 

\section{Starburst Winds}

There is a remarkably simple physical description of what happens when
star formation is rapid in a relatively small volume. The inevitable
outcome is a very strong wind.  Chevalier \& Clegg (1985, hereafter
CC) simplify the situation to the equilibrium solution of mass
injection at a rate $\dot{M}$, with accompanying energy injection,
$\dot{E}$, in a sphere of radius $R$.  Since it turns out that the
wind velocities exceed $1000\kms$, gravity can be ignored in a first
approximation. CC provide a full solution at all radii. Here we are
mainly interested in the asymptotic solution at large radii, where we
cast the CC solution in terms of the terminal wind velocity, $u$ and
the mass injection rate, $\dot{M_w}$,
\begin{equation}
\rho_w = {{\dot{M}_w}\over{4 \pi u r^2}}.
\end{equation}
This wind produces a ram pressure of,
\begin{equation}
P_w \equiv \rho_w u^2 = {{u \dot{M}_w}\over{4 \pi r^2}},
\end{equation}
where $R$ is the size of the region into which the mass and energy are
injected.

As representative numbers we will take $R=10^{21}\cm$, about
$1/3$~kpc, and $\dot{M}_w=10^{27} \gm\psec$, about $15\msun\yr^{-1}$,
approximately the mass injection rate expected during a burst of star
formation of $1500\msun\yr^{-1}$. We follow CC and use
$u=2000 \kms$. In the central region,
\begin{equation}
\rho_w \simeq 0.296 {{\dot{M}_w}\over{ u R^2}},
\end{equation}
or $1.5\times 10^{-24}\gm$, or a number density of about
$1 \cm^{-3}$. 

The cooling time, $t_{cool}= 3kT/(n \Lambda(T))$, at $T=10^8$K where
the cooling rate is about $\Lambda\simeq 3\times 10^{-23}
\cm^3\psec$ is about $4\times 10^7\yr$. The flow time
across the region is only $1.6\times 10^5\yr$, so the hot wind does
not have time to cool. Denser regions in pressure equilibrium will
cool more quickly so that the ISM is unstable and bound to consist of
the hot wind phase and one or more cool phases. Many of the
aspects of this situation are discussed in Ikeuchi \& Norman (1991).

\subsection{Ram Pressure Stripping}

A major objective of this paper is to estimate the ram pressure
stripping of the hot wind on infalling objects.  We calculate the
effects of ram pressure stripping, but note that transport processes,
such as turbulence and heating, could help to increase the rate of gas
removal (\cite{nulsen}).  The calculation proceeds in a series of
steps. First, we derive for our specific case the fairly standard
results that the wind will have a very high momentum flux. The wind
has a sufficiently low density that it will move out before cooling.
The infalling objects are taken to be angular momentum conserving, but
maximally dissipated disks in dark halos, approximated as truncated
Mestel disks. The fractional stripping can be easily estimated for
these objects. The strength of the starburst wind is calculated under
the assumption that star formation is occurring on timescale comparable
to the crossing time of the bulge. Bringing these elements together
gives an expression for the fractional mass of an infalling object
which succeeds in joining the bulge, Equation~\ref{eq:fin}.  

The ridge line of Kormendy's relation is $\mu_B = 3.02 \log{r_0}
+19.74$ B mag arcsec$^{-2}$ (\cite{k77,dejong}), where $\mu_B$ is the
B band surface brightness and $r_0$ is the bulge scale radius in kpc.
For a constant mass-to-light ratio this translates to $\rho\propto
r_0^{-2.2}$. This density radius relation would be slightly weakened
if we allowed for a decrease in mass-to-light for lower luminosity
systems. The implied densities are high enough that bulges are
self-gravitating.  If the rate of infall is, on the average, regulated
by the starburst winds, then the fact that the infall rate is very
insensitive to the mass of the host implies that the characteristic
radius and density of the bulge will scale roughly as $\rho\propto
R^{-2}$, as in Equation~\ref{eq:k}. This relation is roughly the
ridge line of the Kormendy relation.  This is physically easy to
understand. The total bulge star formation rises as the bulge gas
density, $\rho_b,$ to the 3/2 power. For a given accretion rate. a
rise in bulge density will increase the SFR, and hence the outgoing
wind, which temporarily reduces accretion, allowing the gas density to
be reduced. 

The ram pressure rises as $r^{-2}$ with decreasing radius. At $R$, the
outer radius of the star forming volume, the surface density
below which stripping occurs is rises to its maximum,
\begin{equation}
\Sigma {V_c^2\over R} = {{u \dot{M}_w}\over{4 \pi R^2}},
\end{equation}
where $V_c$ is the circular velocity of the incoming gas in its dark
halo.  For the starburst numbers above, $\Sigma=0.16\gm\cm^{-2}$ is
the maximum surface density that can be blown away via ram pressure
alone.  For comparison. the central surface density of a disk galaxy
is about $1 \gm \cm^{-3}$. We conclude that the effects of this wind
would be significant even on disk like that of the Milky Way if it
were completely gaseous.

If we collapse the gas inside an isothermal halo with a velocity
dispersion $\sigma_s$ and an angular momentum parameter $\lambda\simeq
0.05$ to centrifugal equilibrium its surface density increases by
about a factor of $10^2$ over that of the projected isothermal halo,
\begin{equation}
\Sigma(d) \simeq 10^2\Omega_b {\sigma_s^2\over{2d G}},
\end{equation}
where $d$ is the radial co-ordinate in the disk.  For our typical
case, $\Omega_b\simeq 0.01$, $H=100\kmsm$ we find that this
Mestel disk has a 
surface density profile,
\begin{equation}
\Sigma(d) \simeq 0.7\left({\sigma_s\over{100\kms}}\right)^2
	\left({10^{21}\cm\over d}\right) \gm\cm^{-2} .
\end{equation}

The total mass in a Mestel disk of the above form diverges. If we cut
the disk off at the radius of the last orbit that can have come from
the outermost virialized part of the halo we can estimate a total
mass.  The halo is virialized inside approximately $r_{200}$, the
inside of which the mean density is $200\rho_c(z)$, where $\rho_c(z) =
3H(z)^2/(8\pi G)$. For an isothermal sphere, $M(r) = 800/3\pi
\rho_c(z) r_{200}^2 r$.  The isotropic velocity dispersion that
maintains this sphere in equilibrium is, $\sigma_1^2 = 400/3\pi G
\rho_c(z) r_{200}^2$.  With the definition of $\rho_c(z)$ we find that
for the isothermal sphere,
\begin{equation}
r_{200} = {\sqrt{2}\over 10} {\sigma_s\over H(z)}.
\end{equation}
This is a physical (proper) distance.  In this potential the circular
velocity is $V_c = \sqrt{2}\sigma$.

The total gas mass in the disk (assuming that
there are few stars) is $200\rho_c \Omega_b {4\over 3}\pi  r_{200}^3$, or,
\begin{equation}
M_m = \Omega_b {{2\sqrt{2}}\over {10 G H(z)}} \sigma_s^3,
\end{equation}
which is needed to find $\sigma_s$ which is required in the evaluation
of $d/r$ for stripping below, Equation~\ref{eq:fin}.
 
\subsection{Starburst Rates}

Both empirical evidence and theoretical considerations suggest that
the timescale for star formation should be proportional to the local
dynamical timescale (\cite{lh,ken_sfr}). The available data point to a
relation of $t_{sfr} \simeq R/v_b$, where $v_b$ is the local circular
velocity, which in the case of a stellar bulge may be due to
self-gravity, not the dark matter background.  Therefore
$\dot{M}_{SFR} = M v_b/R$, and we will take the wind as $\dot{M}_w =
\epsilon_w \dot{M}_{SFR}$. The mass can be expressed as $M=4\pi/3 \rho
R^3$, where $\rho$ refers to the gas density.

The wind that results from this starburst has a ram pressure,
\begin{equation}
P_w = \epsilon_w v_b u\, \rho {{4\pi}\over 3}{{ R^2}\over{r^2}}.
\end{equation}

\subsection{Self-regulating Starbursts}

The starburst wind will blow away infalling surface densities
smaller than,
\begin{equation}
	\Sigma {V_c^2\over r} <
\epsilon_w v_b u\, \rho {{4\pi}\over 3}{{ R^2}\over{r^2}}.
\end{equation}
In equilibrium this leads to a balance of the bulge which creates
the wind and the and infall,
\begin{equation}
\rho R^2 \simeq \Sigma {3\over{4\pi\epsilon_w u}} {{V_c^2 r_{200}}\over{v_b}}.
\label{eq:k}
\end{equation}
If bulges are to be self-gravitating, then $v_b\gta V_c$, and
generally they are found to have circular velocities comparable to
those of the disk. The important thing to note is that the
quantity $\rho R^2$ is completely determined by the starburst,
whereas the RHS is completely determined by the physics of infall. 

\begin{figure}
\centering
\vspace*{5pt}
\parbox{\textwidth}{\epsfysize=7truecm\epsfbox{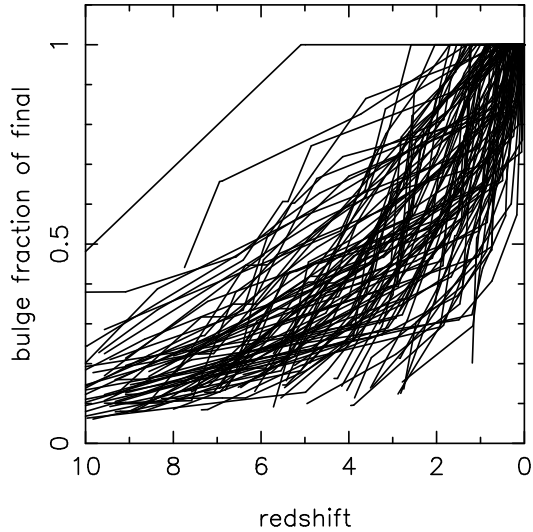}}
\vspace*{5pt}
\caption{Mass redshift history for 100 realizations of bulge building. The
calculation assumes that suitable pre-galactic objects, disks in dark
halos, are available at the beginning of the calculation.}
\label{fig:m_z}
\end{figure}

\begin{figure}
\centering
\vspace*{5pt}
\parbox{\textwidth}{\epsfysize=7truecm\epsfbox{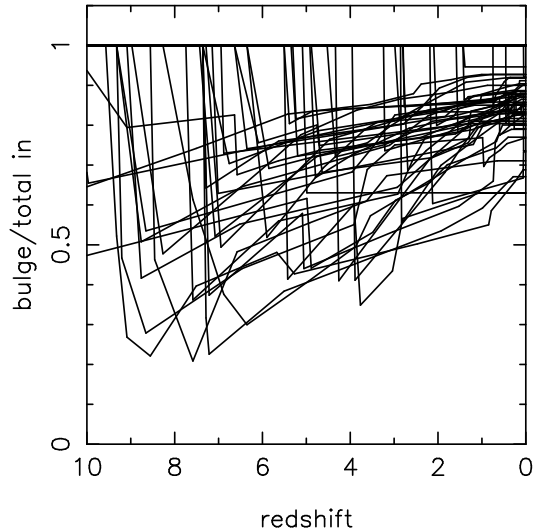}}
\vspace*{5pt}
\caption{Fraction  of infalling gas mass
	successfully accreted versus redshift for the 100
	realizations of bulge building.}
\label{fig:fm_z}
\end{figure}

For Mestel disks derived from satellite halos of $\sigma_s$,
and $\sigma_b^2 =2\pi G  \rho R^2$ are stripped at radii beyond,
\begin{equation}
{d\over r}  > {\sigma_s^2\over\sigma_h^2}
	{3\over{4\epsilon_w}} {v_b\over u} 10^2\Omega_b,
\end{equation}
For typical numbers we take $\sigma_s=\sigma_h/2$ and
$\epsilon_w=10^{-2}$. We find stripping at $d/r>1.3$, which for $r=R$,
and $R\simeq 1/3$~kpc is 99.5\% of the disk mass.

The equation for stripping of Mestel disks can be used to predict
the rate of successful accretion in the presence of a starburst.
We simply multiply $M_m$ with the ratio of $d/r$ for stripping.
The accreted mass is simply 
\begin{equation}
M_{\rm accrete} = {d\over r} {R\over{10 r_{200}}} M_m.
\label{eq:fin}
\end{equation}

\section{Realizations of Bulge Formation Histories}

Combining merging rates and stripped fractions to build bulges with
redshift is now straightforward. To illustrate the model we do a
simple Monte Carlo simulation. We start the simulation at when the
universe is about 0.5 Myr old, prior to significant galaxy creation or
merging.  We adopt an $H_0=65$, $\Omega_M=0.2$, $\Lambda=0$ cosmology,
although the results are not very sensitive to the precise choice of
cosmology. The mergers occur at a rate independent of mass at at a
rate ${\cal R}_n (1+z)^m$. Time is divided into 0.5 Myr intervals and
the probability of an accretion event is the rate per unit time
multiplied with the time interval.  An accretion event occurs if a
[0,1] random number generator produces a number less than this
probability.  The accreted objects begin with a mass drawn from
$\phi(M)\propto M^{\alpha} e^{-M}$, where the normalizing constants
are unity. This therefore assumes that the Mestel disks in their dark
halos are largely present when the calculation is turned on. Clearly
this is not accurate at large redshift, but is arguably a useful
assumption over the redshift zero to about four range
(\cite{steidel}). There is no presumption that the characteristic mass
is as large as that of a full galaxy today, however the characteristic
mass does need to be comparable to a bulge mass, since we find that
the final masses are distributed around unity, the characteristic mass
of the infalling objects.  The wind is assumed to blow uniformly for a
duration of $T_w$ from the previous merger. If a new accretion event
is generated during this interval, then the mass of the incoming
satellite is reduced according to the stripping equation.  The
parameters and their default values are outlined in the following
table.

\begin{table}
\caption{Merger-Wind Simulation Parameters \label{tab:par}}
\begin{center}
\begin{tabular}{lrr}
Parameter & Symbol & Default Value \\ 
Minimum satellite mass & $M_{min}$ & $0.1$ \\ 
Mass function slope	& $\alpha_M$ & $-1.8$ \\ 
Current merger rate & ${\cal R}_n$ & $0.01$ \\ 
Merger-redshift index & $m$	& $1.2$ \\ 
Gas fraction	& $\Omega_{gas}$ & $0.01$ \\ 
Bulge size	& $R_b$	& $10^{21} \cm $ \\ 
Wind duration & $T_w$	& $10^8 \yr$ \\ 
Mass loss efficiency & $\epsilon_w$ & $0.01$ \\ 
\end{tabular}
\end{center}
\end{table}

\begin{figure}
\centering
\vspace*{5pt}
\parbox{\textwidth}{\epsfysize=7truecm\epsfbox{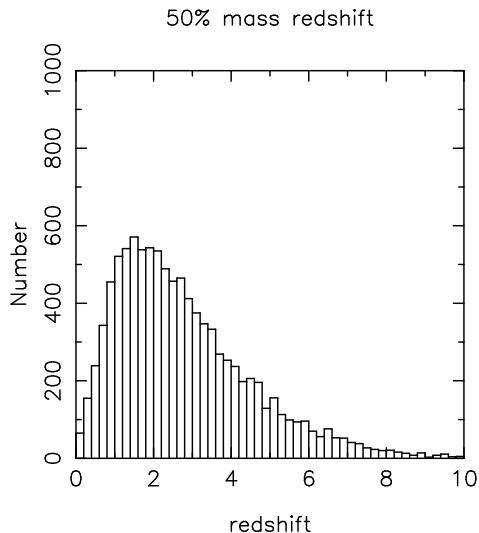}}
\vspace*{5pt}
\caption{Redshift of 50\% mass assembly for 10000 realizations of the
standard model. 
\label{fig:h50}}
\end{figure}

\begin{figure}
\centering
\vspace*{5pt}
\parbox{\textwidth}{\epsfysize=7truecm\epsfbox{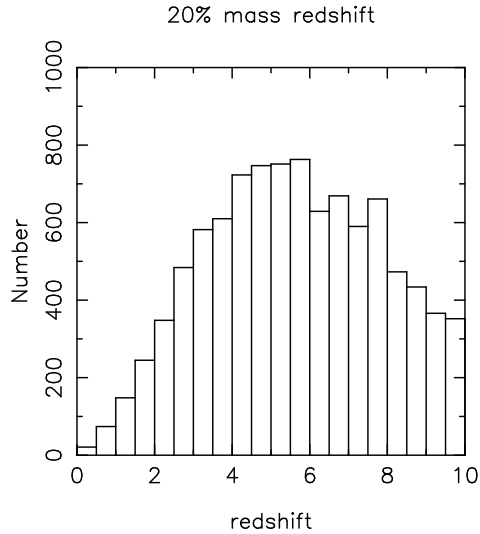}}
\vspace*{5pt}
\caption{Redshift of 20\% mass assembly for 10000 realizations of the
standard model. Comparison with Figure~\ref{fig:h50} emphasizes the
very large spread in redshifts of formation.
\label{fig:h20}}
\end{figure}

In detail there are many (mostly minor) complications that are swept
under the rug here. The model is quite na\"ive in that we assume that
there is a ready supply of gas containing companion galaxies with
roughly a galaxy mass distribution. Observation seems to support this
as being true from redshift zero to about four, which covers most of
the activity here.  Mergers wouldn't do much at all if $\alpha
\gta-1.2$, as is observed at low redshift. We have chosen a steeper
$\alpha$ to both mimic the increase in gas content with decreasing
mass, and, to take into account that $\alpha$ does become steep in the
redshift 3--4 range (\cite{steidel}). The gas content of galaxies
should likely vary with redshift, whereas we have simply taken them to
be all gas. Likely the pairwise velocity dispersion decreases somewhat
with increasing redshift, which will diminish the fraction of low
redshift pairs that merge. Overall the effect of all of these things
would be to decrease the importance of merging at below redshift
one. The natural tendency of the model is to have little activity at
low redshift anyway, so the basic character of the results should not
depend on these simplifications. In any case, the main purpose of this
``toy'' star formation history is to examine the basic viability of
the model, not to fine tune the parameters.

Overall, we find several interesting results. First, the mass buildup
predicted by this simple model seems to be very roughly in accord with
the requirements of ``classical'' bulges. Figure~\ref{fig:h50} 
shows that About 10\% are more than 50\% formed at
redshift 5. But, about 3\% are only half formed at redshift
0.5. Although the median time of half assembly is a reassuring
redshift of about two, there is a tremendous spread of formation
times. The redshifts of 20\% assembly are shown in Figure~\ref{fig:h20}.

\begin{figure}
\centering
\vspace*{5pt}
\parbox{\textwidth}{\hbox{\epsfysize=6truecm\epsfbox{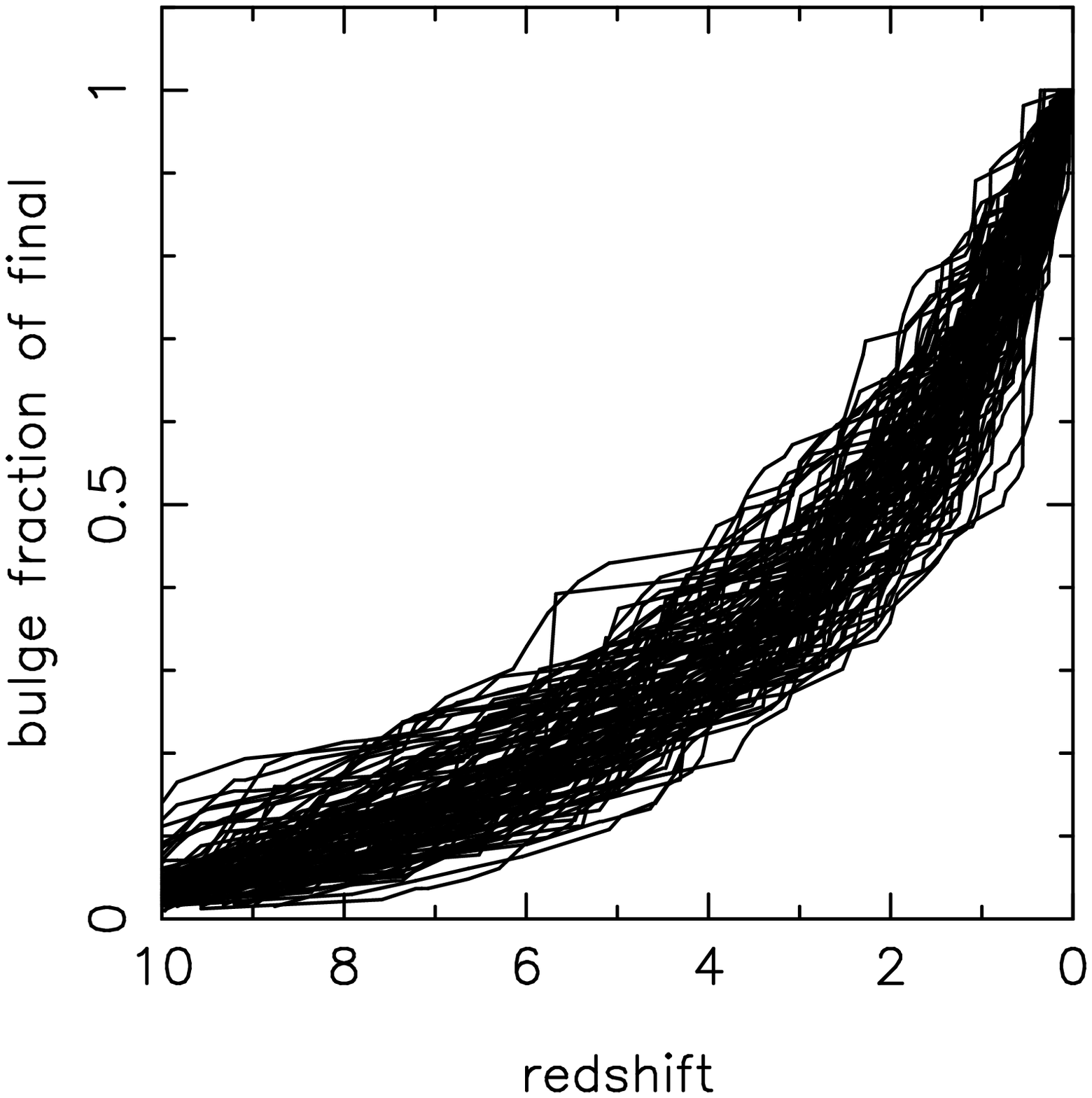}~~
	\epsfysize=6truecm\epsfbox{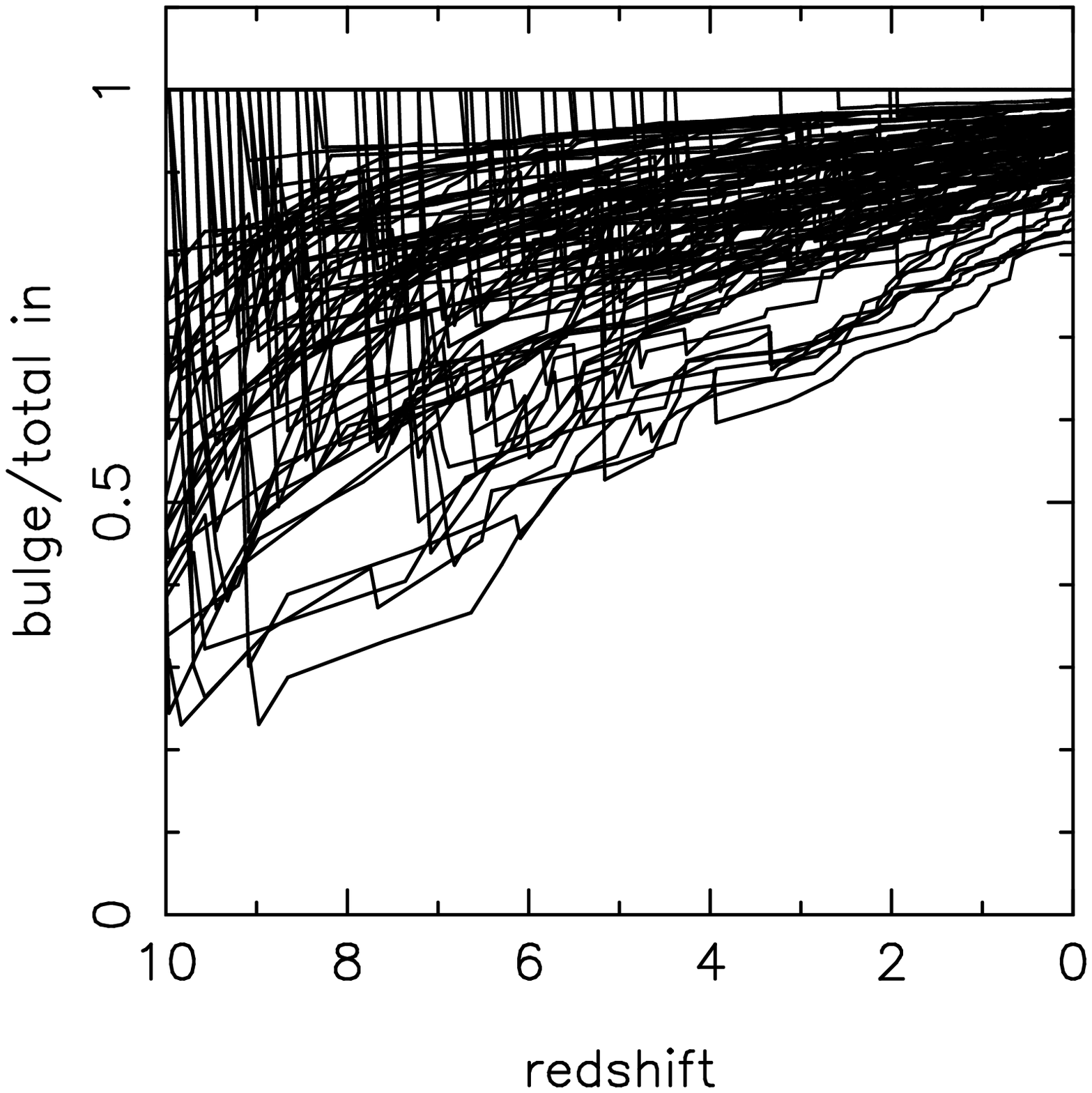}}}
\vspace*{5pt}
\caption{The assembly of a bulge built from pre-galactic objects with
a minimum mass of $0.03M_\ast$, rather than the standard $0.1M_\ast$
and wind lifetime of $10^7 \yr$.  The left panel is the analog of
Figure~\ref{fig:m_z} and the right panel is the analog of
Figure~\ref{fig:fm_z}. The assembly here is smoother and more affected by
the starburst winds. Although there is less dispersion in formation
times, the median 20\% and 50\% time are not significantly different
from the standard model.
\label{fig:sm}}
\end{figure}

The standard wind has a lifetime of $10^8$ years, which in many cases
limits infall to about 2/3 of what it would normally be,
Figure~\ref{fig:fm_z}.  The limitation of infall would help
drive bulges towards the Kormendy relations. Winds are even more
effective if the smallest mass to be merged into a bulge is reduced to
3\% of $M_\ast$, rather than our standard $0.1M_\ast$. This is shown
in Figure~\ref{fig:sm}.  The basic merger rate is the same, so the
redshifts of assembly are not greatly altered.  However because the
bulge is being built of more, smaller, units, the buildup has less
dispersion in time.

\section{Conclusions}

The bulge formation history is predicted here using the observed
density of nearby (gas-rich) galaxies with masses comparable to the
eventual bulges. Further, we have argued that starburst winds will
have a significant effect on the accreting gas. A straightforward
assessment of our results is that the merger history appears to be
roughly in accord with what is known about star formation histories
and bulge ages. An additional step, not taken here, is to use these
assembly histories to predict the color distribution of bulges as a
function of redshift. These then become a simple but powerful test of
the model. The attraction of the merger model is that it is based on
observations, which now provide a very basic measure over the redshift
zero to four range, although the masses of the incoming objects are
not well quantified at the highest redshifts.  Beyond redshift four,
this model is likely becoming less reliable, since the assumption that
the pregalactic fragments are in place has no basis in observation and
even in a low density Universe, the Press-Schechter approach would
indicate that the halos in which these reside are becoming less
numerous. Furthermore the gas content probably evolves with redshift.
An attraction of this approach is that the need for additional
physical parameters can be driven by observation.

The relevance for bulges of the strong starburst winds that we have
advocated is far less clear at this stage.  The attraction of the idea
is that it promotes the development of the Kormendy relations in a
self-gravitating system. The tests of this model will require fairly
detailed observations of bulges at high redshifts, particularly
concentrating on the heavily reddened objects in which starbursts
generally occur. The presence of starburst winds and their affects can
potentially be detected in optical emission lines, X-ray observations
of hot gas (\eg \cite{heck_wind}) and ultimately in {\em resolved}
observations of the molecular gas (\eg \cite{frayer}) which is known
to exist.

\begin{acknowledgments}
This research was supported by NSERC of Canada. I thank the Carnegie
Observatories, Pasadena, for their hospitality during the time when
this work was initiated.
\end{acknowledgments}

\def\apj{ApJ}
\def\apjl{ApJL}
\def\apjs{ApJS}
\def\aj{AJ}
\def\mnras{MNRAS}
\def\araa{ARAA}
\def\aap{A\&Ap}
\def\aaps{A\&ApS}

\clearpage

\end{document}